\documentclass[preprint,1p]{elsarticle}

\usepackage{xcolor}

\usepackage{lineno}

\usepackage{booktabs}
\usepackage{graphicx}
\usepackage{caption}
\usepackage{url}
\usepackage{subcaption}
\usepackage{amsmath}
\usepackage{float}
\usepackage{imakeidx}
\usepackage[utf8]{inputenc}

\usepackage{natbib}
\journal{Nuclear Inst. and Methods in Physics Research, A}

\begin{document}

\begin{frontmatter}
\title{Iterative-Bayesian unfolding of isotopic cosmic-ray fluxes measured by AMS-02}

\author[1]{E. F. Bueno\corref{cor1}}
\ead{e.ferronato.bueno@rug.nl}
\author[2]{F. Barão}
\ead{barao@lip.pt}
\author[1]{M. Vecchi}
\ead{m.vecchi@rug.nl}

\cortext[cor1]{Corresponding author}

\address[1]{Kapteyn Astronomical Institute, University of Groningen,Landleven 12, 9747 AD, Groningen, the Netherlands}
\address[2]{Laboratório de Instrumentação e Física Experimental de Partículas (LIP), 1649-003 Lisboa, Portugal}

\begin{abstract}

The measurement of the isotopic composition of cosmic rays (CRs) provides essential insights the understanding of the origin and propagation of these particles, namely the CR source spectra, the propagation processes and the galactic halo size. The Alpha Magnetic Spectrometer (AMS-02), a CR detector operating aboard the International Space Station since May 2011, has the capability of performing these measurements due to its precise determination of the velocity provided by its Time of Flight (TOF) and Ring Imaging Cherenkov (RICH) detector. The correct interpretation of the data requires the measurements to be deconvoluted from the instrumental effects. The unique design of AMS-02, with more than one subdetector being used to measure the same flux, requires a novel approach to unfold the measured fluxes. In this work, we describe an iterative-Bayesian unfolding method applied in the context of isotopic flux measurements in AMS-02. The accuracy of the method is assessed using a simulated flux based on previous measurements and a full detector response function. We introduce a non-parametric regularization method for the detector response functions, as well as a single, smooth prior flux covering the full range of measurements from both detectors, TOF and RICH. In addition, the estimation of the errors and a discussion about the performance of the method are also shown, demonstrating that the method is fast and reliable, allowing for the recovery of the true fluxes in the full energy range.
\end{abstract}

\end{frontmatter}

\section{Introduction}
\label{intro}

 The elemental abundance of light isotopes such as $^{2}$H, $^{6}$Li and $^{10}$Be in the Solar System is low if compared to their abundance in CRs, which indicates these are particles from secondary origin: they are produced during the interaction of primary CRs, those produced and accelerated at the sources, with the insterstellar medium (ISM) \citep{gaisser}. Therefore, the measurement of the flux of such isotopes provides essential information about the CR source composition and the propagation processes in our Galaxy \cite{coste}. Moreover, radioactive species such as $^{10}$Be also provide important constraints on the galactic halo size \cite{hayakawa}. Magnetic spectrometers such as AMS-01\citep{d_ams01}, PAMELA\citep{PAMELA_d} and IMAX\cite{IMAX92} have measured the fluxes of some of these isotopes. In such detectors, the mass identification is performed by combining the measurements of the rigidity (momentum per unit charge, $R=pc/Ze$) and the velocity, measured either directly or through energy deposition, $\beta$. Both measurements can be combined in a single observable, the mass, which is given by the equation

\begin{equation}
    m = \frac{RZ}{\beta\gamma}
\end{equation}

\noindent where $Z$ is the magnitude of the charge and $\gamma$ is the Lorentz factor.
The measurements before AMS-02, reaching at most a few GeV/n, are limited by detector velocity and rigidity resolutions. In AMS-02, isotopes are identified either in the velocity and rigidity phase space \citep{ams-helium-isotopes}, or through the use of mass templates in events binned in velocity \cite{efb}.

Additionally to particle identification, one of the main challenges in particle physics experiments is the reconstruction of the true properties of the particles. The finite resolution of the detector, energy losses and other instrumental effects lead to biasing and smearing of quantities such as velocity and rigidity. In the case of the measurement of isotopic spectra, where they are generally obtained as a function of the kinetic energy per nucleon or rigidity, one must face this problem to recover the spectra of the particles before the particles interact with the detector. Generally speaking, the true value of a variable in a given bin, $x_{t}^j$, is related to its measured value, $x_{m}^i$, through the response function of the detector, $R(x_m^i|x_t^j)$:
\begin{equation}
    x_m^i = \sum_j R(x_{m}^i|x_{t}^j) x_{t}^j
\end{equation}

\noindent The problem is then to invert the previous equation, obtaining the values of $x_t^j$. Naively inverting this expression leads to numerical instability; hence other methods have been developed to obtain \textit{estimators} of $x_t$. This estimation procedure is frequently called unfolding and is widely used in the field of astroparticle physics, especially in direct measurements of CR fluxes, where several unfolding techniques have been applied in different measurements \citep{PAMELA_d}\citep{IMAX92}\citep{ams-helium-isotopes}. In this paper, we will focus on the application of an iterative approach, based on Bayes' theorem, similar to the works of D'Agostini \citep{dagostini} and Mülthei and Schorr \citep{schorr}, in the context of the measurement of isotopic fluxes with AMS-02, where the events are counted as a function of the velocity measured by two different subdetectors, TOF and RICH.

This paper is organized as follows: in section 2 we given and overview of AMS-02, describing the TOF and RICH in more detail. In section 3 we recall the ingredients for the calculation of the flux, followed by the development of the flux unfolding expressions based on Bayes' theorem. In section 4 we describe in detail the application of the proposed unfolding technique in mock fluxes in AMS-02, from the construction of the response function to the error estimation. We conclude in section 5.

\section{The Alpha Magnetic Spectrometer}

The Alpha Magnetic Spectrometer (AMS-02) is a particle physics detector operating aboard the International Space Station since May 2011, having collected over 200 billion events so far. AMS-02 consists of different sub-detectors\citep{AMSPhysReport}:
the nine-layer silicon tracker, combined with the 0.15 T permanent magnet, measures the magnitude and sign of the charge, and the rigidity ($R = pc/Ze$) of the particle; the Transition Radiator Detector, used to separate leptons from hadrons; the Time of Flight (TOF), responsible for measuring the velocity of the particle, also acting as the main trigger of AMS-02; the Ring Imaging Cherenkov Detector (RICH), which measures the magnitude of the charge, as well as the velocity; the Anti-Coincidence Counter (ACC), that rejects particles with high-incidence angle; and the Electromagnetic Calorimeter (ECAL), located right below the RICH,  used to measure the energy of the particle, as well as to separate leptons from hadrons. 

AMS-02 can measure the fluxes of isotopes in CRs in a wide energy range because of the accuracy and complementarity of the velocity measurements provided by the TOF and the RICH. The TOF consists of two pairs of layers of scintillator planes, one above and the other below the magnet. The time taken for a particle to travel between the pairs of scintillators while traversing the magnetic field, $\Delta t$, is registered. The length of the path taken by the particle, $L$, is obtained with the track reconstruction made by the silicon tracker. In possession of these quantities, the velocity $\beta$ is simply $\beta = \Delta t/cL$, and has a resolution $\Delta \beta / \beta^{2} \approx 4\% $ for single-charged particles and $\beta \approx 1$ \citep{TOFPerformance}. The RICH is a proximity-focusing detector which measures $\beta$ using the Cherenkov effect. It includes a radiator plane in which particles travelling with a velocity above the speed of light in the medium emit electromagnetic radiation peaked in the UV region \citep{amsrich}. The radiation emitted is detected below in a plane of  $4 \times 4$ pixelized photomultipliers. The PMTs of the detection plane are arranged to be outside the ECAL acceptance, thus reducing the amount of matter traversed by the particles before they reach the calorimeter. Considering this arrangement, the radiator plane is made of two materials; its center consists of sodium fluoride (RICH-NaF), with a refraction index $n = 1.33$ and threshold of emission $\beta = 0.75$. The NaF produces broad Cherenkov rings which avoid the ECAL hole, increasing the acceptance of the detector. Surrounding this material there is the aerogel (RICH-AGL) with a refraction index $n = 1.05$ and threshold of emission $\beta = 0.96$. the velocity resolution in the NaF and in the AGL are $\Delta \beta / \beta \approx 0.35\%$ and $\Delta \beta / \beta \approx 0.12\%$, respectively,  for particles with $Z=1$ and $\beta \approx 1$ \citep{AMSPhysReport}.

Monte Carlo (MC) simulations of protons were also used in this study. These simulations were produced by the AMS collaboration through a dedicated software based on the GEANT4 package \cite{GEANT4}. The software simulates the interactions of particles with AMS material. It produces detector responses, which are then used to reconstruct the desired event properties in the same way as in data.

\section{Flux unfolding}

The flux of a particle can be expressed as

\begin{equation}
\label{eq:flux}
    \Phi_m(x) = \frac{N(x)}{ A(x) \varepsilon(x) T(x) \Delta x }
\end{equation}

\noindent where $N(x)$ is the number of events in the accumulated number of events in a finite interval of $x$, $\varepsilon(x)$ the selection efficiency, $A(x)$ the geometric acceptance of the detector, $T(x)$ is the exposure time for particles above the geomagnetic cutoff; and $\Delta x$ the interval width. In the flux expression, $N(x)$ is an observable and therefore is a function of the measured variable, $x_m$; $A$ is a constant factor; $\varepsilon(x)$ is estimated from Monte Carlo simulations. Hence it is obtained as a function of the true variable, $x_t$; $T(x)$ is, strictly speaking, dependent on the measured variable, but due to the use of a detector resolution safety factor, as done in \citep{ams_pflux}, this dependence vanishes. 

Since $N(x)$ includes the instrumental effects, the obtention of the true number of events requires the application of an unfolding method. The following expression relates the measured number of events in a given bin, $N(x_m^i)$, to its true value, $N(x_t^i)$

\begin{equation}
    N(x_m^i) =  \sum_{j} R(x_m^i|x_t^j)\varepsilon(x_t^j)\eta_0(x_t^j)N(x_t^j)
    \label{eq:nmeas}
\end{equation}

\noindent where the factor $R(x_m^i|x_t^j)$ is the response function of the detector, and $\eta_0(x_t^i)$ is an efficiency translating the probability of a true value being observed inside the detection range, that is, $\eta_0(x_t^j) = \sum_i R(x_m^i|x_t^j)$. The efficiency $\varepsilon(x_t^i)$, although estimated from simulated events as a function of the true variable, $x_t$, needs to be corrected for eventual discrepancies between data and simulations on the efficiencies of selection criteria, that is

\begin{equation}
    \varepsilon(x_t) \rightarrow \varepsilon(x_t) \underbrace{\prod_{k} \frac{\varepsilon_{Data}(x_m^k)}{\varepsilon_{MC}(x_m^k)}}_{\delta(x_m)}
\end{equation}

\noindent where $\varepsilon_{MC}$ and $\varepsilon_{Data}$ are, respectively, efficiencies as estimated from simulations and experimental data for the set ($k$) of data selection criteria. This ensures that the obtained efficiency values agree with those from the actual experiment. Hence, equation \ref{eq:nmeas} becomes

\begin{equation}
    \frac{N(x_m^i)}{\delta(x_m^i)} =  \sum_{j} R(x_m^i|x_t^j)\varepsilon(x_t^j)\eta_0(x_t^j)N(x_t^j)
    \label{eq:nmeaseff}
\end{equation}


The goal of the unfolding procedure is to obtain the true flux, $\Phi_{t}(x_t)$, which corresponds to the incident flux purged from instrumental effects. To this end, the true number of events, $N(x_t)$ must be obtained. The inversion of equation \ref{eq:nmeaseff} is intrinsically unstable and therefore cannot be done directly. Using Bayes' theorem, one can estimate the probability of the true value being $x_t$ in the bin $i$, when $x_m$ was measured

\begin{equation}
    \label{eq:posterior}
    P(x_t^i |x_m^j) = \frac{P(x_m^j |x_t^i) P(x_t^i)}{\sum_k P(x_m^j |x_t^k) P(x_t^k)}
\end{equation}

\noindent where he probability $P(x_m^j |x_t^i)$ represents the \textit{likelihood} that $x_m$ in bin $j$ was produced by $x_t$ in bin $i$, while $P(x_t^i)$ represents the \textit{prior} knowledge of $x_t$. As a result, the probability $P(x_t^i |x_m^i)$ is the \textit{posterior}, or the updated knowledge of $x_t$. In the context of this problem, the likelihood is the response function of the detector $R(x_m|x_t)$, and the prior is the previous knowledge about $N(x_t)$. With an estimate of the response function, and with some prior knowledge of $N(x_t)$, it is possible to use equation \ref{eq:posterior} to estimate the true value from the measured one. This is done in the following equation, where the true number of events in the bin $i$ is derived from the accumulated number of events, the efficiency corrections and the posterior probability.

\begin{equation}
\label{eq:nhat}
    \hat{N}_{t}(x_t^i) = \frac{1}{\varepsilon(x_t^i) \eta_0(x_t^i)}\sum_j \frac{N_{m}(x_m^j)}{\delta(x_m^j)} P(x_t^i |x_m^j)
\end{equation}

\noindent

\noindent $\hat{N}_t(x_t^i)$ represents the \textit{estimator} of $N(x_t^i)$, and is called the unfolded number of events. The terms outside the sum represent the efficiency corrections. Since the posterior is naturally dependent on the prior, a convergent iterative process has to be established, with $\hat{N}_t(x_t^i)$ being used to define a prior for the next iteration. In each step, the prior is regularized by means of a fit to the data. The convergence of the method is evaluated through the use of the $\xi$ estimator:

\begin{equation}
\label{eq:chi2}
    \xi_{k+1, k} = \sum_i\left(\frac{\hat{N}_t^{k+1}(x_t^i) - \hat{N}_t^{k}(x_t^i)}{\hat{N}_t^{k}(x_t^i)}\right)^2
\end{equation}

\noindent where $k$ is the number of the iteration.

In the next section, the method will be applied in the context of the AMS-02 experiment using a simulated flux that mimics the actual conditions of the experiment.

\section{Application to mock AMS-02 fluxes}

\begin{figure}
    \centering
    \includegraphics[width=\linewidth]{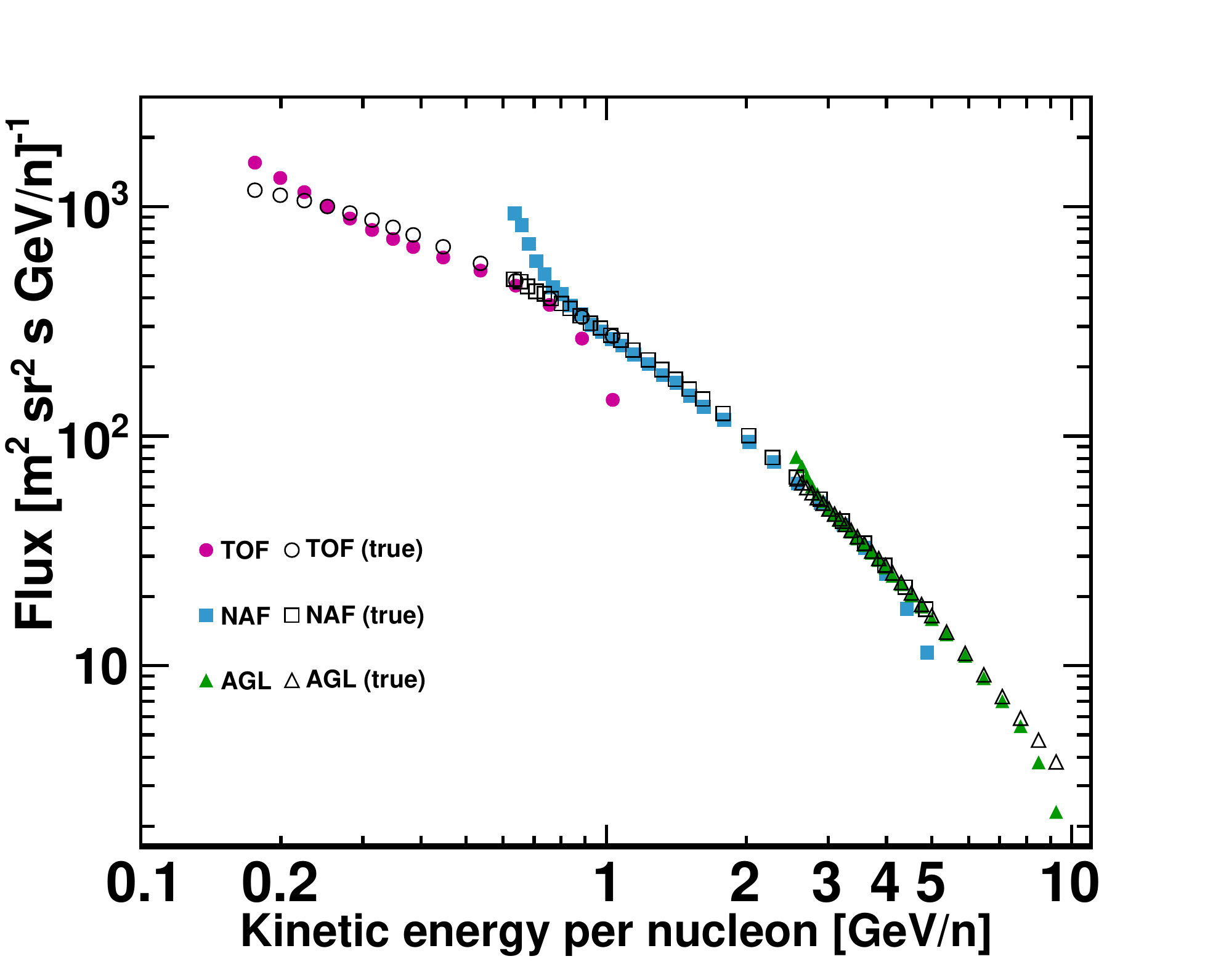}
    \caption{True (empty) and measured (solid) fluxes versus the kinetic energy per nucleon for the TOF (circles), RICH-NaF (squares) and RICH-AGL (triangles).}
    \label{fig:sim_flux}
\end{figure}

In order to show the method at work in the context of single-charged isotopes flux measurements in AMS-02, it has been applied to mock fluxes. Considering the precise measurements of the velocity provided by the TOF and the RICH detectors, the fluxes are measured as a function of the velocity in three overlapping ranges: $0.5 < \beta < 0.9$ in the TOF; $0.78 < \beta < 0.99$ in the RICH-NaF; and $0.96 < \beta < 0.996$ in the RICH-AGL. The events were generated following the AMS-02 proton spectrum \cite{ams_pflux} parametrization given in \citep{pinching}. They were then smeared and biased according to the three detector response functions. In order to reproduce the effect of data selection efficiencies, generated events were weighted according to the selection efficiencies in every range. Figure \ref{fig:sim_flux} shows the measured and generated fluxes as a function of kinetic energy per nucleon, $E_k = (\gamma - 1)m_p$, where $\gamma$ is the Lorentz factor and $m_p$ is the proton mass. The noticeable differences between the true and measured fluxes emphasize the necessity of using a method to correct the measurements for the detector response.

In the following subsections, we describe the construction of the posterior probability for the unfolding procedure.

\subsection{Detector response function regularization}

\begin{figure}
    \centering
     \includegraphics[scale=0.75]{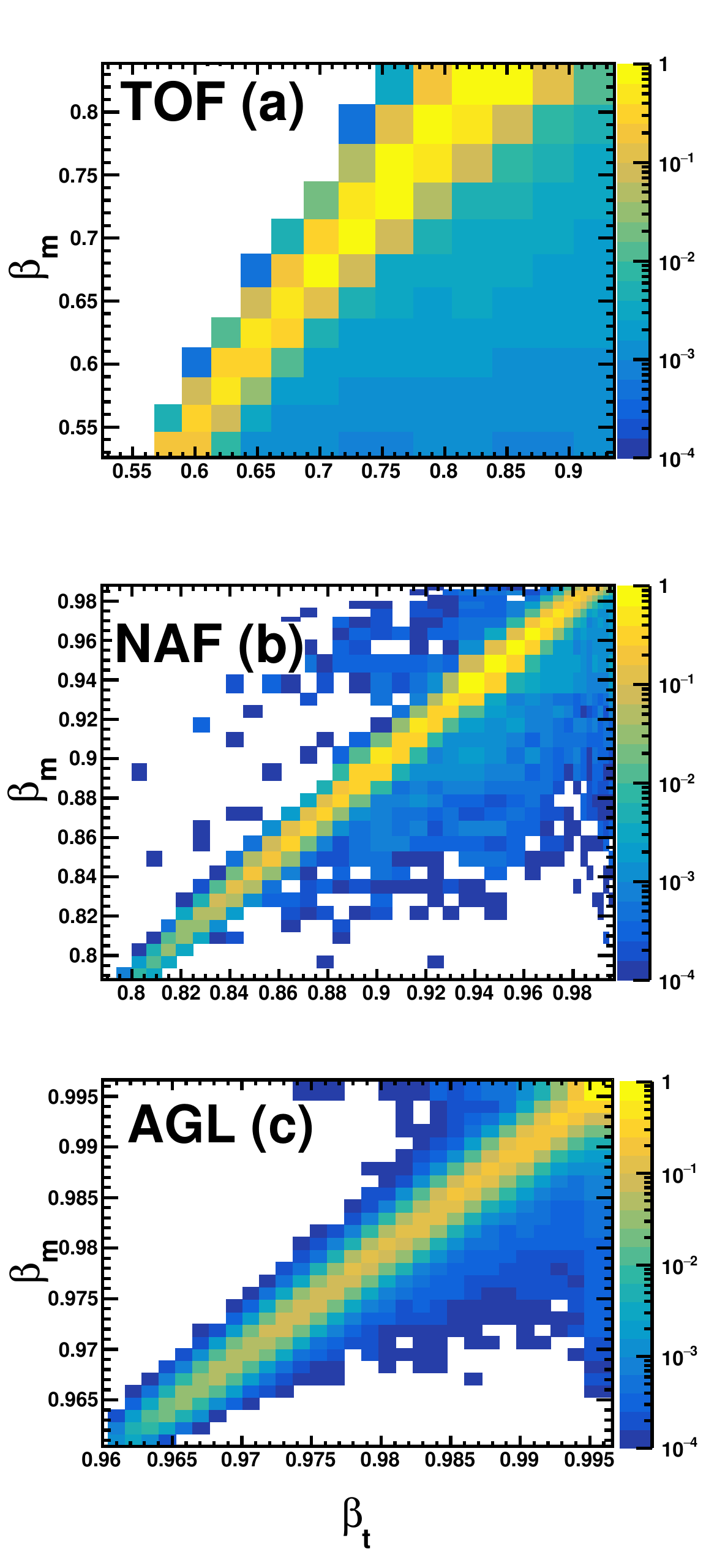}
    \caption{Migration matrices of the TOF (a), RICH-NaF (b) and RICH-AGL (c). See text for discussion.}
    \label{fig:mig_matrix}
\end{figure}

The detector response functions, often referred to as migration matrices, are obtained from the Monte Carlo simulations of the detector. As in the isotopic flux measurements, the events are counted in bins of velocity (or, equivalently, kinetic energy per nucleon). These matrices correspond to binned distributions of $\beta_{t}$, the velocity generated in the simulation, versus $\beta_{m}$, the velocity as reconstructed in the detector. Figure \ref{fig:mig_matrix} shows these response functions for the TOF (a), RICH-NaF (b) and RICH-AGL (c). The TOF has a larger velocity bias at low energies due to the energy loss undergone by the particles. Both RICH-NaF and RICH-AGL are less penalized by this effect.

\begin{figure}
    \centering
    \includegraphics[width=\linewidth]{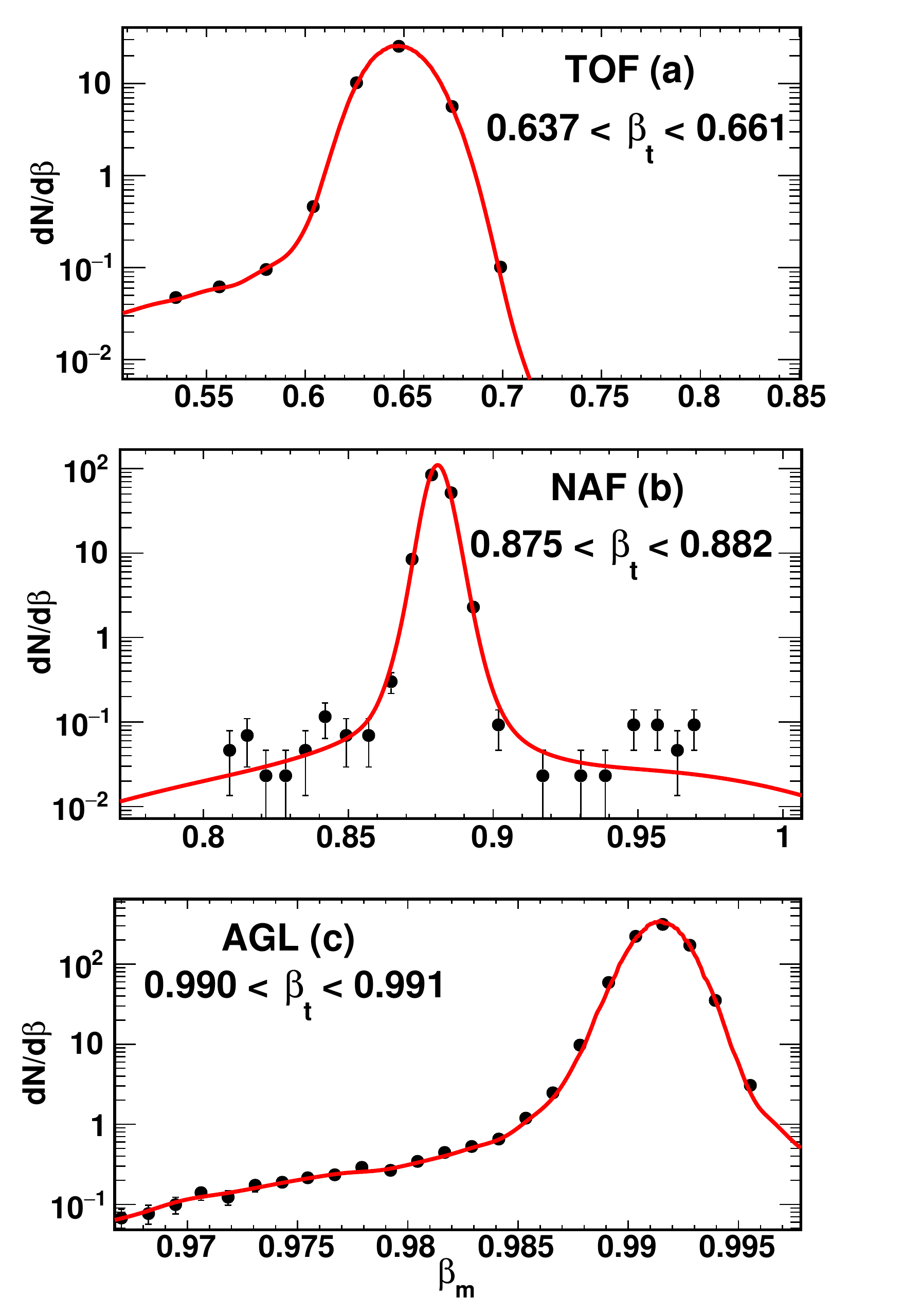}
    \caption{The resolution functions for the TOF (a), RICH-NaF (b) and RICH-AGL (c) in different true velocity bins. See text for discussion.}
    \label{fig:matrix_proj}
\end{figure}

Given that the simulations have a finite number of events, regularising the matrices is a crucial step to avoiding fluctuations caused by the statistical uncertainties. To this end, the one-dimensional projections used for calculating the likelihood are regularized to obtain smooth distributions. Figure \ref{fig:matrix_proj} shows examples of these projections for the TOF, RICH-NaF, and RICH-AGL in panels (a), (b), and (c), respectively, for three velocity intervals. The $\beta_{m}$ distribution in the TOF is composed of a Gaussian core and a tail corresponding to events whose velocity reconstruction was heavily affected by the energy loss. For the two RICH regions, tails on the measured velocity are visible for two main reasons: the low number of Cherenkov photons in the ring produced by single-charged particles and the presence of additional hits in the RICH PMT plane caused by secondary particles from protons interacting with the AMS-02 material. In order to have a smooth description of the detector response functions, we proceeded to their regularization by using the kernel-density-estimation technique (KDE)\citep{kdes}. This non-parametric method eases the smoothing of features that would need complex parametrizations to be described, albeit at a higher computational cost.

The obtained parametrizations, $f(\beta_m | \beta_t)$, were then used to calculate the probabilities $P(\beta_m^j | \beta_t^i)$ by integrating them over the bins of measured velocity, $\beta_m^j$, for a given bin of true velocity, $\beta_t^i$:

\begin{equation}
    P(\beta_m^j | \beta_t^i) = \int_{\beta_m^j} f(\beta_m | \beta_t) d\beta_m
\end{equation}

\noindent This process is repeated for every bin, allowing for the construction of a smooth migration matrix.

\subsection{Prior construction}

\begin{figure}[!h]
    \centering
    \includegraphics[width=\linewidth]{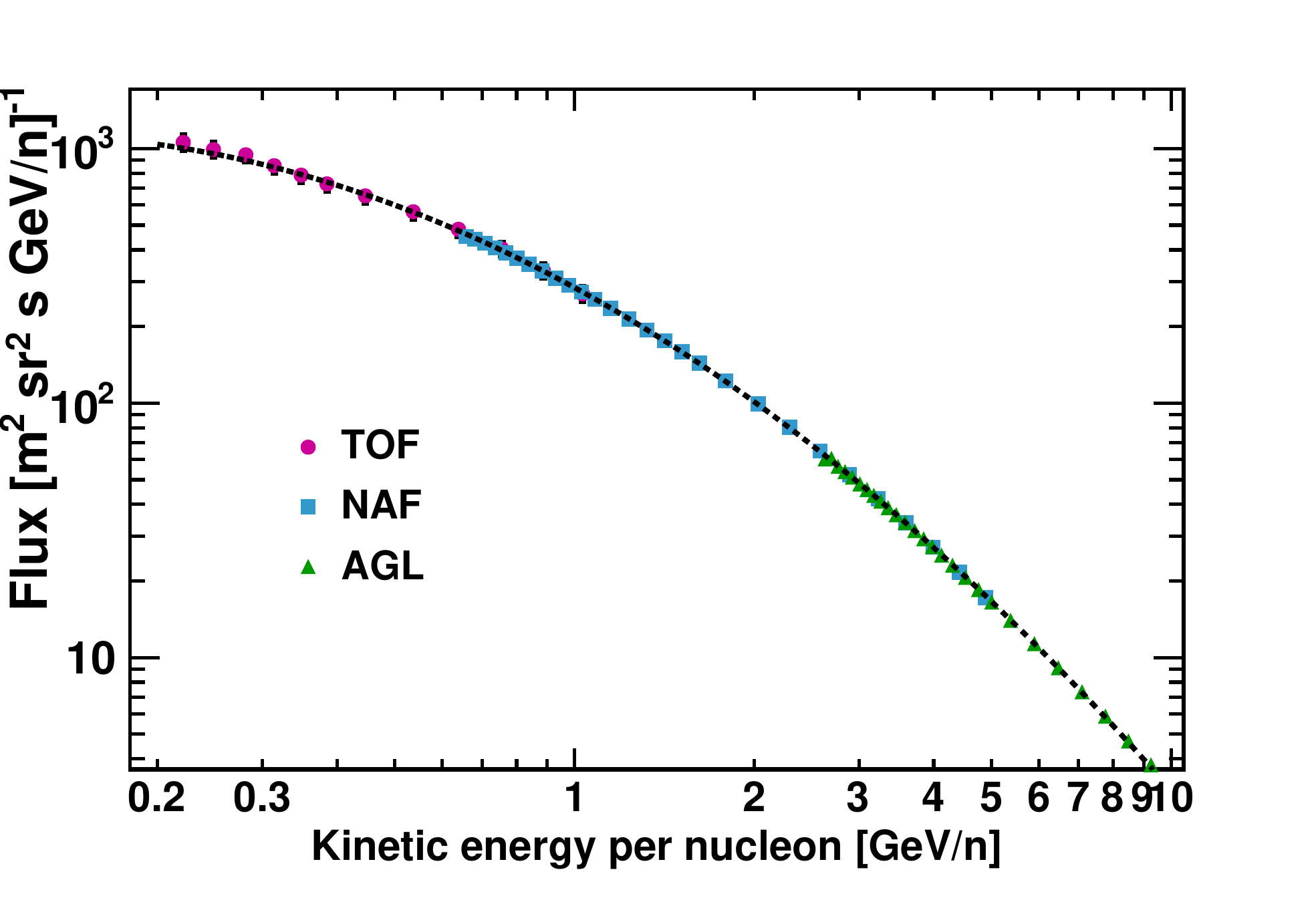}
    \caption{Model (black dashed line) adjusted on the results of the second iteration of the TOF (pink circles), RICH-NaF (blue squares) and RICH-AGL (green triangles).}
    \label{fig:prior}
\end{figure}

The prior, representing the degree of belief in the true variable, is a critical component in calculating the posterior defined in equation \ref{eq:posterior}. Due to the iterative nature of this method, the choice first prior cannot affect the result. Hence, a natural choice for the initial prior is a flat shape. After every iteration, the unfolded number of events, $\hat{N}(E_k(\beta)_t)$, is calculated using equation \ref{eq:nhat} for every detector range, allowing for the calculation of the corresponding flux according to the equation

\begin{equation}
\label{eq:unf_flux}
     \Phi(E_k)_t = \frac{\hat{N}(E_k)_t}{ A(E_k)_t \  T(E_k)_t \ \Delta E_k }
\end{equation}
These fluxes are regularized through a fit of the model described in \citep{pinching}, performed on the three ranges simultaneously as shown in figure \ref{fig:prior}. This parametrization represents an updated degree of belief in the flux as a function of the true variable and is used to obtain the prior for the next iteration through equation \ref{eq:unf_flux}: the updated flux is used together with the acceptance, exposure time and bin width to compute the new prior probability of $N(E_k)_t$. The advantage of such an approach is that it describes the data well, providing a smooth description of the flux, even in transition zones between the ranges, where resolution and threshold effects may cause the points to fluctuate, as seen in figure \ref{fig:prior}. This guarantees that the prior for the next iteration is a smooth, continuous function. In addition, it ensures that the unfolded fluxes match in the transition zones between the detectors. It prevents the fluctuations that may come from the unfolding procedure from propagating to the next iteration, ensuring the method converges.

\subsection{Error estimation}

The calculation of the unfolded number of events depends directly on the detector response function, as well as on the number of measured events. These quantities have errors associated with them, which are propagated to the result of the unfolding.
In order to estimate the contributions from each of these quantities, a bootstrapping method was implemented. This method consists of generating $n$ randomized sets of data based on the original one and performing the unfolding procedure in all sets, resulting in a distribution of unfolding results with $n$ entries per bin. This allows for calculating the covariance matrices, which contain the total uncertainties in each bin in its diagonal. Considering iterative nature of the procedure, where the results of one iteration are used as the prior for the next, the propagation of the uncertainty is done by sampling the prior.

Assuming that the uncertainties on the number of measured events and on the migration matrices are uncorrelated, they are treated separately. For the contribution from the number of measured events, $10^3$ sets of measurements were generated by sampling new values $N(E(\beta)_m^i)$ in every bin according to a Gaussian distribution. That is, the number of events in each bin $i$ of each new set, $N'(E(\beta)_m^i)$, is obtained by sampling a Gaussian distribution $G$ centered at $N(E(\beta)_m^i)$, with RMS given by \\ $\sqrt{N(E(\beta)_m^i)}$, that is

\begin{equation}
\label{eq:sampling}
    N'(E(\beta)_m^i) = G(N(E(\beta)_m^i); \sigma_{N})
\end{equation}

\noindent The unfolding procedure was applied to the complete set of measurements, yielding a distribution of unfolded counts. Using the distribution of the results, the covariance matrix of the unfolded number of events was obtained.

After every iteration, the prior is updated with the previously obtained unfolded counts. We associate with the prior the uncertainty obtained in the previous iteration and propagate it by sampling the unfolded counts according to their uncertainty, analogously to what is stated in equation \ref{eq:sampling}.




A similar exercise was performed to obtain the errors coming from the uncertainties on the migration matrices. This time, the measured number of events was kept fixed, and $10^3$ migration matrices were generated according to

\begin{equation}
     P(E(\beta)_m^j | E(\beta)_t^i) = G(P(E(\beta)_m^j | E(\beta)_t^i) ; \sigma_{P})
\end{equation}

\noindent where G is a Gaussian distribution centered at the original $ P(E(\beta)_m^j | E(\beta)_t^i)$, and width given by $\sigma_P$, that represents the uncertainty on each migration matrix bin. Analogously to the previous case, the unfolding is performed with the $10^3$ different matrices, resulting in a covariance matrix unfolded counts per velocity range, from where the uncertainties were obtained.

\begin{figure}[!h]
    \centering
    \includegraphics[width=\linewidth]{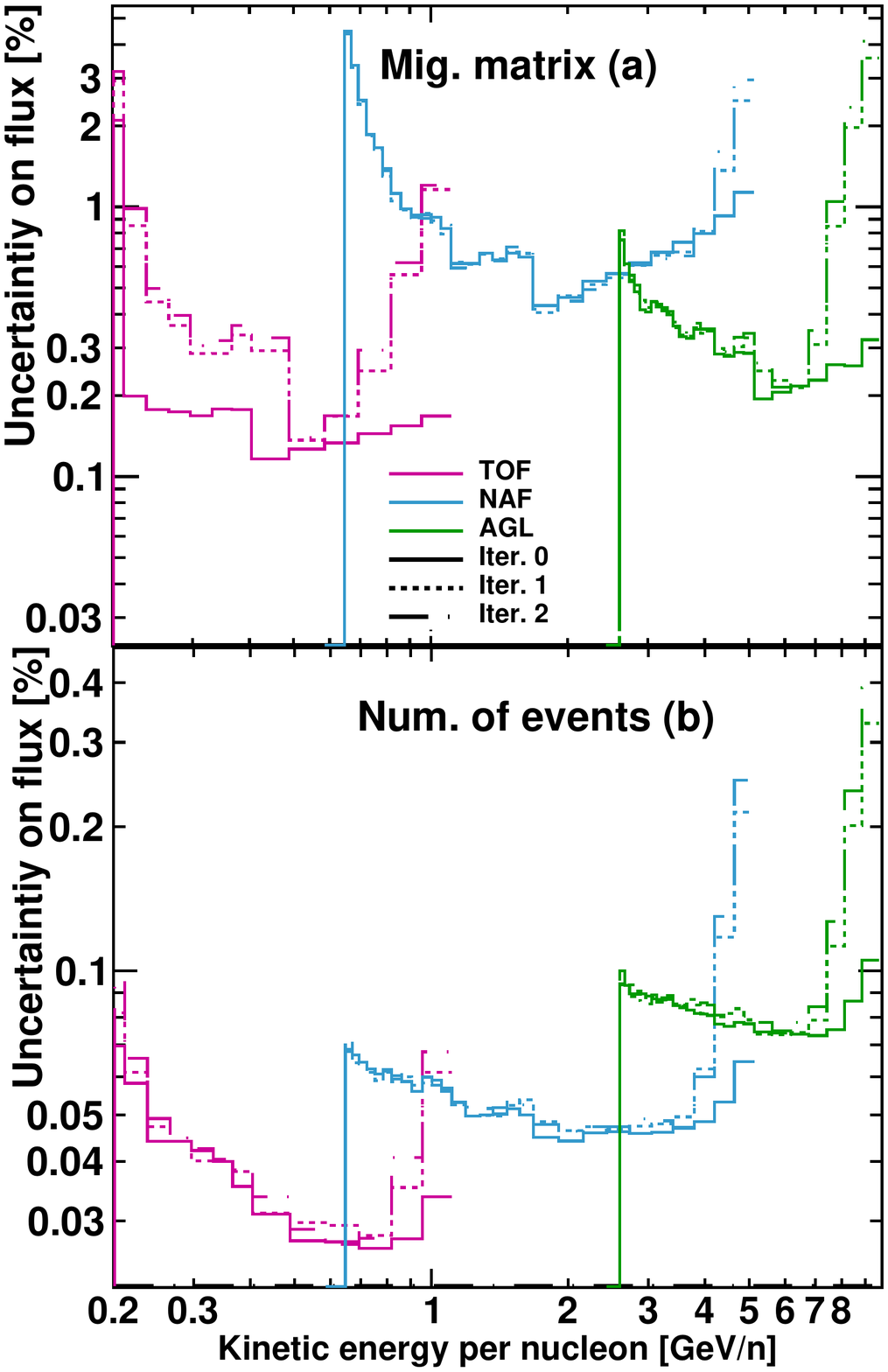}
    \caption{Relative error on unfolded flux for the migration matrix contribution (top) and on the number of measured events (bottom). Pink lines represent the TOF, blue the NAF, and green the AGL. Solid lines are the errors from iteration 0 (flat prior), while dotted and dot-dashed correspond to the errors for iterations 1 and 2, respectively.}
    \label{fig:rel_errors}
\end{figure}

The relative errors for both contributions, coming from the diagonal of the covariance matrices,  are shown for the different ranges and iterations in figure \ref{fig:rel_errors}. Panel (a) shows the uncertainty coming from the migration matrices. In all cases, the errors are larger in the extremes of each range. The explanation for the increase at the end is the same for the three subdetectors: at the upper edges of each velocity range, the velocity resolution degrades, hence increasing the migration of events. Therefore, fluctuations in the migration matrices lead to fluctuations in the result. The increase at the beginning of each range is due to different physical phenomena: in the TOF, it is due to the energy loss, which produces a strong bias in the response function at the lowest energies. Hence any response fluctuations in that region of the matrix produce significant changes in the unfolded results. The errors on the RICH follow a similar trend, but in this case this behavior is explained by the Cherenkov emission kinetic energy per nucleon threshold. Events close to the threshold have a smaller probability of being reconstructed. Hence the migration matrix is less populated in this region. As a result, the dispersion of the unfolding results increases, increasing the error. The same behavior is seen in panel (b), where the uncertainty coming from the number of events is shown.

The errors from both contributions increase progressively with the number of iterations, as expected from the error propagation done via the prior sampling. The error increases faster from iteration 0 to 1 than from 1 to 2. This is due to the change in the prior: while iteration 0 has a prior which is flat in the entire energy range, iterations 1 and 2 use a model which is fitted to the results, leading to larger uncertainties.

Once both contributions are estimated, the total error of the unfolding procedure is given by the sum in quadrature of the uncertainties from the migration matrix and the number of events.

\subsection{Number of iterations}

\begin{figure}[!h]
    \centering
    \includegraphics[width=\linewidth]{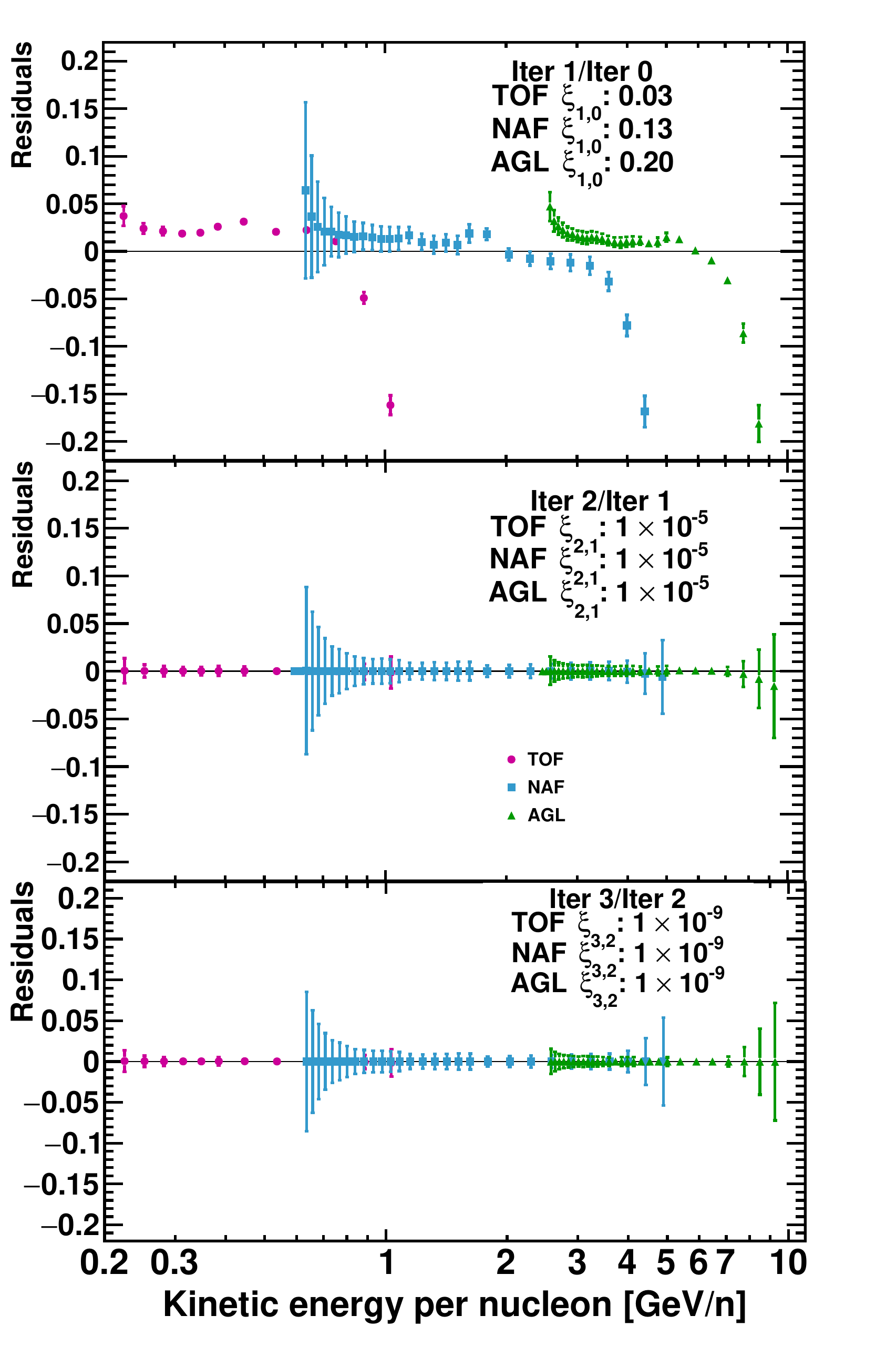}
    \caption{Residuals and  $\xi$ calculated for consecutive iterations in the three velocity ranges. The top panel shows the residuals between iterations 0 and 1, the mid panel shows between 1 and 2, and the bottom shows between 3 and 2. The different ranges are represented by different markers and colors: pink circles (TOF), blue squares (NAF) and green triangles (AGL). See text for discussion.}
    \label{fig:iters}
\end{figure}

Considering the iterative nature of this method, the number of iterations is an important parameter. Too few iterations lead to a result far from the true value, while a large number of iterations leads to a larger error due to the error propagation. Therefore, one must stop iterating as soon as the differences between two consecutive iterations are small enough. Figure \ref{fig:iters} shows the residuals between consecutive iterations, defined as

\begin{equation}
    R^{k+1,k}_i = \frac{N^{k+1}_i - N^{k}_i}{N^{k+1}_i}
\end{equation}

 where $k$ is the number of the iteration and $i$ is the bin index. In addition, the value of $\xi$ defined in equation \ref{eq:chi2} is shown for every range in each of the panels. The largest changes are between iteration 0 and 1 due to the use of a flat prior in iteration 0. Iterations 1 and 2 already show signs of convergence, with the residuals displaying small fluctuations in the last points of the AGL range. The bottom panel shows that iterations 3 and 2 yield essentially the same results, with $\xi_{3,2} \approx 10^{-8}$ meaning that it is enough to stop at iteration 2.

The similarity between iterations 3 and 2 highlights the importance of regularizing the prior in each iteration. The spline fit removes eventual fluctuation that appear in the calculations and ensures convergence, whereas the methods with no regularization have oscillations due to positive feedback of these fluctuations \citep{dagostini}.

\subsection{Comparison with true flux}

\begin{figure}[!h]
    \centering
    \includegraphics[width=\linewidth]{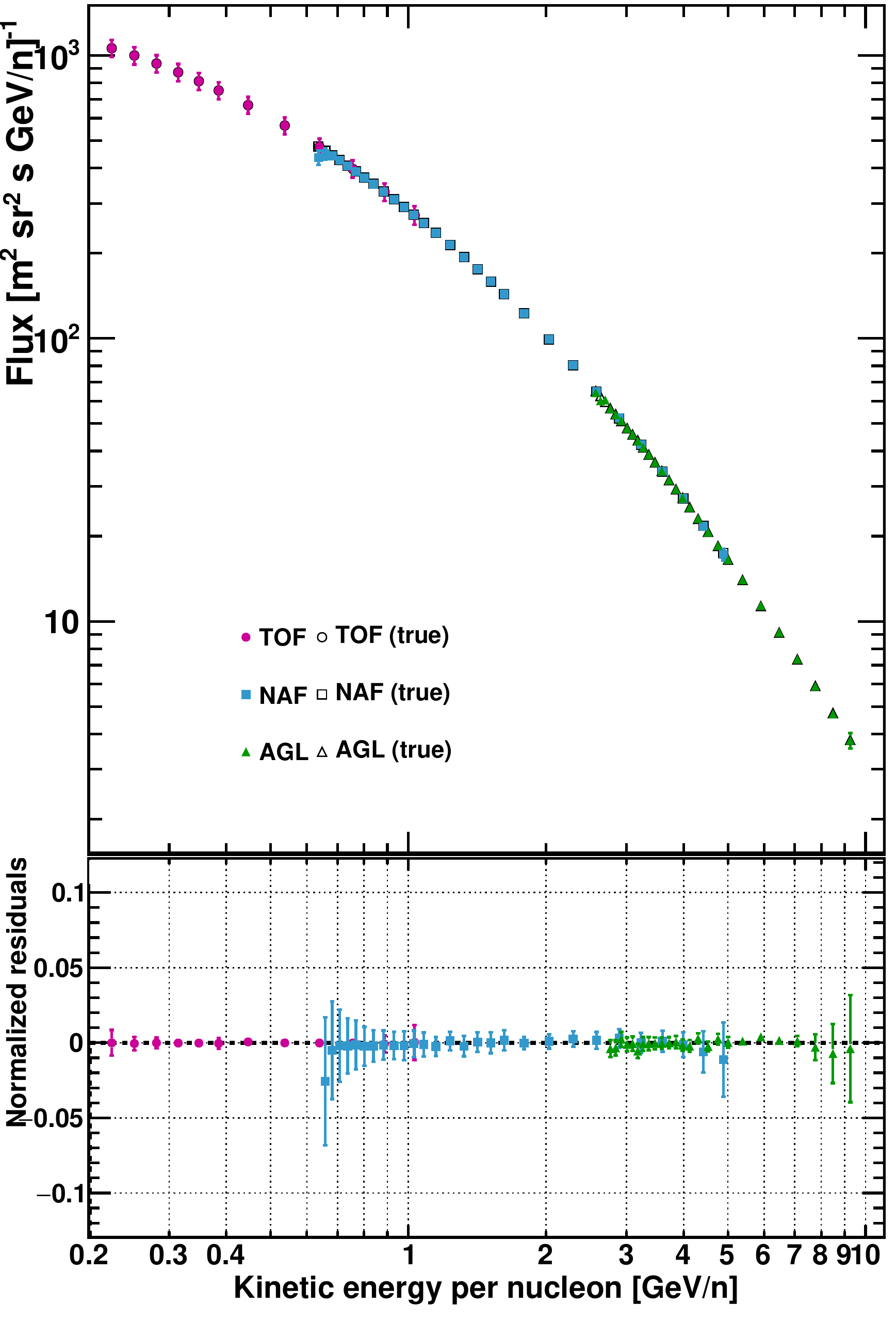}
    \caption{True and unfolded flux after three iterations (top) and the normalized residuals (bottom). See text for discussion.}
    \label{fig:results}
\end{figure}

 The top panel of figure \ref{fig:results} shows the unfolded flux after three iterations, calculated with equation \ref{eq:unf_flux}, for the different ranges, while in the bottom, the residuals comparing the results with the true fluxes are shown. Generally, the agreement between the unfolded and true values is very good, being below $1\%$ in the entire measurement range. Minor fluctuations are seen at the beginning of the RICH-NaF and RICH-AGL ranges. These are related to steep drops in the number of measured events due to the efficiency of the detector, which makes the true distribution harder to obtain in these regions. In the case of the RICH the efficiencies drop rapidly below 0.8 and 3 GeV/n, respectively, due to the emission threshold of Cherenkov radiation. Still, the method can recover the true distribution quite well in these regions. There is also a good agreement between the detectors in the overlapping regions.

\section{Conclusion}

The study of the isotopic composition of CRs will provide important information about the propagation of CRs in our own Galaxy, as well as about the CR source composition and the Galactic halo size. These measurements are intrinsically challenging due to several factors: large backgrounds, as in the case of single-charged isotopes, and small mass difference, as is the case of Be isotopes. As a consequence, the available measurements are limited to the kinetic energy per nucleon range from sub-GeV/n to a few GeV/n. The AMS-02 experiment, with its precise measurements of the rigidity and the velocity, especially provided by the Cherenkov detector, has the capability of measuring the fluxes of isotopes reaching energies up to 10 GeV/n. The correct interpretation of these data depends on correcting the results for the detector response. The unique design of AMS-02, with three different measurements of the velocity, requires special treatment. In this work, we applied an iterative unfolding method based on Bayes' theorem in the context of the isotopic flux measurements in AMS-02. The general mathematical aspects of the method were discussed, emphasizing the importance of data/simulation and efficiency corrections. In order to demonstrate the applicability and performance of the method, a mock flux was generated according to the measured proton spectrum, which was then smeared and biased according to the response function of the different sub-detectors included in the analysis. Regularization was also introduced in two stages of the method. The response function was regularized with KDEs, which could readily describe the non-Gaussian features of the measurements. In addition, the unfolding of the flux in the three different ranges was connected via a common prior, obtained through the fit of a model using all the available points. This fit leads to the smoothing of features in the flux, such as the transition between ranges, and prevents fluctuations from leading the method to calculation to diverge. The error, estimated through a bootstrapping procedure, has shown strong dependence on the migration matrix, growing rapidly in regions with high bias or low statistics. Finally, the results have shown that only a few iterations of this procedure are enough to recover the true spectrum within the uncertainties, enabling the true flux to be estimated up to 10 GeV/n. The fast convergence of this method also makes it suitable for studying the time-variability of isotopic fluxes, where a large number of flux unfolding calculations are necessary.

\section*{CRediT authorship contribution statement}
\textbf{E. F. Bueno:} Conceptualization, Methodology, Software, Data Curation, Formal analysis, Visualization, Writing - Original draft preparation. \textbf{F. Barao:} Conceptualization, Methodology, Software, Writing - Review $\&$ Editing. \textbf{M. Vecchi:} Writing - Review $\&$ Editing,  Funding acquisition, Visualization, Supervision.

\section*{Declaration of competing interest}
The authors declare that they have no known competing financial interests or personal relationships that could have appeared to influence the work reported in this paper.

\section*{Acknowledgements}
This publication is part of the project "Statistical methods applied to cosmic ray anti-deuteron searches with the AMS-02 experiment" with project number 040.11.723 of the research programme Bezoekersbeurs 2019 VW, which is financed by the Dutch Research Council (NWO). 
This publication is also part of the project "Title Cosmic ray antideuterons as a probe for new physics" with project number OCENW.KLEIN.387/11680 of the research programme Grant Open Competition Domain Science - M OC ENW KLEIN which is (partly) financed by the Dutch Research Council (NWO). We are grateful to Portuguese FCT for the financial support through grant CERN/FIS-PAR/0007/2021.

\bibliography{AMS}

\end{document}